\documentclass{aa}   
\usepackage[normalem]{ulem}
\usepackage{xcolor}
\usepackage{booktabs}
\usepackage{graphicx}
\usepackage{txfonts}
\usepackage{lipsum}
\usepackage{subcaption}         % necessary for continued figures, example in section 3
\usepackage{lscape}             % to rotate a single page table, example in appendix.
\usepackage{placeins}

\begin{document}

   \title{Tracing the main belt origins of S-complex near-Earth objects: From space mission targets to meteorite sources}

\author{
    P.N.~Simon\inst{\ref{lam}}       
    \and 
    P.~Vernazza\inst{\ref{lam}}                                     
    \and
    M.~Marsset\inst{\ref{eso}}
    \and 
    M. Brož\inst{\ref{prag}} 
    \and 
    R.P.~Binzel\inst{\ref{mit}}                 
    \and 
    F.~DeMeo\inst{\ref{mit}}
    \and
    C.A.~Thomas\inst{\ref{nau}}
    \and 
    L.~Jorda\inst{\ref{lam}}          
}
           
\institute{
      %---- Vernazza
     Aix-Marseille University, CNRS, CNES, LAM, Institut Origines, 38 rue Frederic Joliot Curie, 13388 Marseille, France
     \label{lam}
     \and %---- Marsset
    European Southern Observatory (ESO), Karl-Schwarzschild-Strasse 2, 85748 Garching bei München, Germany
     \label{eso}
     \and
    Charles University, Faculty of Mathematics and Physics, Institute of Astronomy, V Holešovičkách 2, CZ-18200, Czech Republic       
     \label{prag}
     \and %---- Binzel, DeMeo
     Department of Earth, Atmospheric and Planetary Sciences, MIT, 77 Massachusetts Avenue, Cambridge, MA 02139, USA
     \label{mit}
     \and %---- Thomas
     Department of Astronomy and Planetary Science, Northern Arizona University P.O. Box 6010, Flagstaff, AZ 86011, USA
     \label{nau}
}

  \abstract
   {Space missions to near-Earth objects (NEOs) provide remarkable opportunities to study and characterise their physical and compositional properties in detail. Interpreting these results in a broader context requires knowledge of their provenance, and therefore of their source family in the main asteroid belt.}
   {In this work, we aim to identify the family sources of six S-complex NEOs visited or soon to be visited by space missions. A secondary objective is to investigate the size-dependent orbital behaviour of the broader S-complex NEO population using the MITHNEOS spectroscopic dataset. } 
   {We combine two complementary approaches: near-infrared spectral fitting of NEO spectra with ordinary chondrite (OC) analogues reddened by an empirical space-weathering model, using reduced $\chi^2$ minimisation, and cross-validation with probabilistic dynamical maps of the main NEO source families. }
   {We identify the closest OC type and the most probable source family of the six targets: LL6–7 chondrites and the Flora family for Eros, Itokawa, Apophis, and Torifune; LL3 chondrites and the Nysa family for Toutatis; L chondrites and the Massalia family for Didymos-Dimorphos. When in situ observations are available (Itokawa and Eros), the classifications are consistent. Applied to the broader S-complex NEO population dataset, the spectral classification and orbital properties distributions are broadly consistent with dynamical simulation models. When accounting for observational biases, we find that at small diameters ($\leq$100 m), the fractions of H- and L-type objects increase at the expense of LL6–7 types, with their low-inclination orbital properties being consistent with origins in the Massalia (L) and Koronis (H) families.}
   {Combining spectral fitting with OC analogues and dynamical probability maps provides a robust method of identifying the source families of individual NEOs. Below $D\sim$100 m, the compositional distribution of S-complex NEOs begins to approach that of meteorite falls. This represents direct observational evidence, based on spectroscopic measurements of S-complex NEOs, of the emergence of the source regions of OC meteorites. }

\titlerunning{Main belt origins of S-complex NEOs}
\maketitle
\nolinenumbers 
\section{Introduction}

In recent years, substantial progress has been achieved in identifying the source regions of meteorites and near-Earth objects (NEOs) by integrating spectroscopic data for both populations with numerical simulations of the collisional and dynamical evolution of the major asteroid families in the main asteroid belt \citep{Broz2024_a, Broz2024_b, Marsset2024}.
It has been shown that recent ($\leq$40 Myr) major asteroid breakups tend to dominate the meteoroid flux at metre sizes (Massalia2: L chondrites; Karin+Koronis2: H chondrites; Veritas: CM chondrites), whereas families with the largest number of $D\geq$1 km objects located near the two most favourable dynamical routes (the $\nu_6$ and 3:1 resonances) tend to dominate the NEO flux at kilometre sizes (e.g. Phocaea: H-like NEOs; Juno: L-like NEOs; Flora: LL-like NEOs). For LL chondrites, these associations have been recently refined by \citet{marsset2026nysafamilymainsource}, with the Nysa family identified as the main source of unequilibrated LL3 chondrites, whereas the Flora family appears to be the principal source of equilibrated LL4–7 chondrites.

While significant progress has been achieved for large-scale populations (meteorites, NEOs), the origin of NEOs visited by space missions remains debated and ambiguous. However, understanding their provenance is of prime importance for contextualising the scientific results of these missions. 

In this paper, we consider six S-complex NEO targets explored by recent, ongoing, or upcoming missions. Some have already been visited: (433) Eros, explored in 2000 by NASA's {NEAR Shoemaker} mission \citep{Veverka2001}; (25143) Itokawa, visited in 2005 by JAXA's sample-return mission {Hayabusa} \citep{Fujiwara_2006}; (4179) Toutatis, observed during a flyby by the CNSA {Chang'e 2} mission in 2012 \citep{Huang2013}; and (98943) Torifune, the target of a flyby by JAXA's {Hayabusa2\#} mission in July 2026 \citep{TSUDA2025254}. The remaining objects are targets of ongoing or upcoming missions: (65803) Didymos and its satellite Dimorphos, impacted by NASA's {DART} mission in 2022 \citep{Michel_2022} and targeted by ESA's \textit{Hera} rendezvous mission in 2026 \citep{MICHEL20182261}; and (99942) Apophis, the target of both ESA's RAMSES \citep{Michel2025_ramses} and NASA's OSIRIS-APEX \citep{Nolan2025} missions. This study is particularly timely, as these missions will soon provide new constraints on the physical and compositional properties of these bodies.

We estimated the source families of all S-complex NEOs visited by space missions by combining the analysis of their spectral properties with a comparison of their orbital properties to dynamical probability maps of the main families for kilometre-sized NEOs \citep{Broz2024_a, Broz2024_b, Broz_2026}. For consistency and to validate our identified source families, we also reanalysed the full set of available near-infrared (NIR) spectra of S-complex NEOs \citep{Marsset_2022, Marsset2024} as well as newly available data from the MIT-Hawaii Near-Earth Object Spectroscopic Survey (MITHNEOS). 

\section{Spectroscopic datasets and reduction}
The NIR (0.8–2.5 $\mu$m) spectral dataset was primarily acquired by MITHNEOS using the SpeX instrument \citep{Rayner_2003} on the 3-metre NASA Infrared Telescope Facility (IRTF) at Mauna Kea, Hawaii. We initially selected spectra of S-complex near-Earth asteroids (NEAs) or their derivatives according to their classification from the SsODNet database \citep{SsODNet}. Spectra flagged as low quality or lacking typical S-complex spectral features were excluded, resulting in a sample of 813 spectra corresponding to 651 objects. The spectra were already reduced following the procedure described in the MITHNEOS survey \citep{BINZEL201941}. Supplementary spectra were included for three of the six NEAs specifically targeted in this study: Apophis \citep{Reddy_2018}, Torifune \citep{Geem_2023, DEMEO2009160}, and multiple spectra of the Didymos-Dimorphos system \citep{Polishook_2023}, acquired both before and after the DART impact. A complete list of spectra used for each target is provided in Appendix~\ref{app:spectra_table}.
Given the diverse and not always fully documented reduction pipelines for these supplementary datasets, we reassessed their uncertainties using conservative error estimates. These were defined as the maximum of the provided uncertainty estimates, local spectral variability, and a minimum uncertainty floor of 1$\%$ of the reflectance flux. Additional uncertainty margins were applied in wavelength regions affected by strong telluric absorption. 

For the mean spectra of H and L ordinary chondrites (OCs), we adopted the ones from \cite{Vernazza_2014}. For LL OCs, we distinguished between petrologic subtypes (unequilibrated LL3 versus equilibrated LL6–7), following the results of \cite{marsset2026nysafamilymainsource}.

\section{Methods: Spectral analysis}
\label{sec:methods}

\begin{figure}
    \centering
    \includegraphics[width=\linewidth]{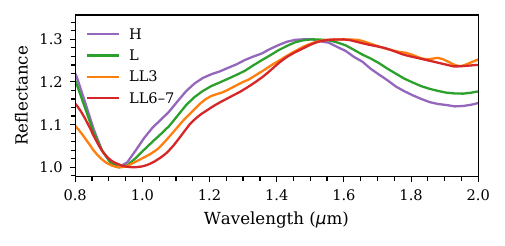}
    \caption{Average NIR reflectance spectra of the four OC subclasses considered in this study: H, L, LL3, and LL6–7. The spectra were normalised at 0.93 $\mu$m and scaled to a common Band I contrast to emphasise differences in band shape, following the fitting procedure described in Sect. \ref{sec:methods}. Across the OC subclasses, Band I becomes progressively broader and its absorption minimum shifts to longer wavelengths with increasing olivine content, from H (narrowest) to LL6–7 (broadest).}
    \label{fig:OC_spectra}
\end{figure}

\begin{figure}[!h]
    \centering
    \includegraphics[width=\linewidth]{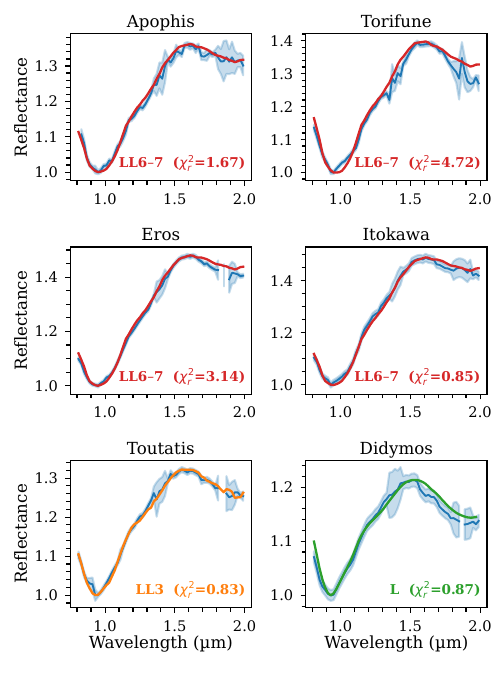}
    \caption{Example NIR reflectance spectra for each of the six targets (in blue), with their estimated uncertainties, fitted by the model described in Sect. \ref{sec:methods}. For each spectrum, only the best-fit analogue and its $\chi^2_r$ value are shown.}
    \label{fig:chi2_fitting_6NEAs}
\end{figure}

To determine the meteoritic analogue of each body, we adopted the same approach as detailed in \cite{Marsset2024}. Specifically, asteroid spectra were compared with the four OC subclasses considered in this study (H, L, LL3, and LL6–7), whose spectral differences are illustrated in Fig. \ref{fig:OC_spectra}. Each chondrite spectrum was first normalised to its minimum reflectance and then fitted to the asteroid spectrum using the empirical space-weathering model of \cite{Brunetto_2006}, with additional constraints requiring the fit to pass through both the Band I minimum and the right shoulder of Band I. In this way, both the meteorite and the asteroid share the same band depth. Finally, reduced $\chi^2$ values were computed for each chondrite, and the lowest value identifies the most probable analogue. For illustration, Fig. \ref{fig:chi2_fitting_6NEAs} shows the best-fitting OC spectrum for one representative spectrum of each of the six targets.
For asteroids with multiple spectra, as is the case for all targets of this study except Itokawa, we chose not to average them, as they may correspond to different observing conditions (e.g. phase angle or surface region). Instead, each spectrum was analysed independently, and the resulting $\chi^2_r$ values were examined as a distribution to assess the dispersion in fit quality and the consistency of the compositional classification.

However, the spectral and compositional overlap between adjacent chondrite types in the case of L and LL3, and between LL3 and LL6–7, is significant. Therefore, we extended the fitting to two additional mean spectra computed as the average of the respective boundary pairs: (L+LL3)/2 and (LL3+LL6–7)/2. If (L+LL3)/2 yields the best fit, the object is classified as L, and if (LL3+LL6–7)/2 is preferred, it is classified as LL6–7.
The motivation for this convention is detailed in Appendix \ref{app:mineralo_distrib}. Nevertheless, these choices are deliberate and may marginally underestimate the contribution of LL3-like objects, and consequently that of the Nysa family.
 

\section{Results}

\subsection{Spectral classification of mission targets}

\begin{figure}[!h]
    \centering
    \includegraphics[width=\columnwidth]{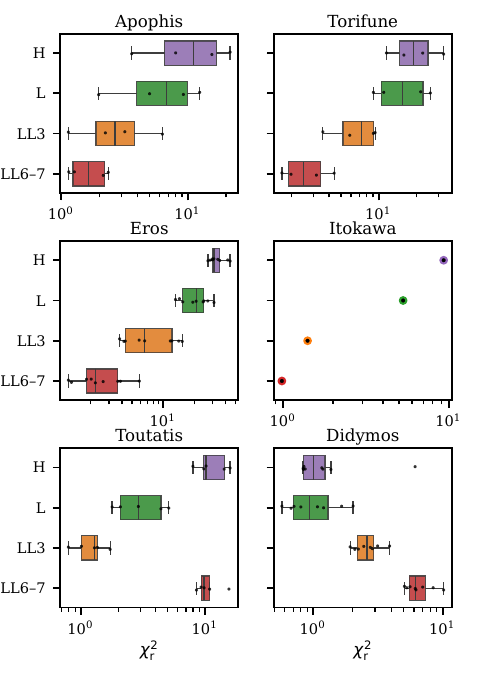}

    \caption{Distribution of the $\chi^2_{\mathrm{r}}$ values obtained for each targeted asteroid and each chondrite analogue. Each box represents the interquartile range (25–75$\%$) of the $\chi^2_{r}$ values from multiple spectra of the same asteroid. The median is indicated by the central line, and whiskers show the range excluding outliers.}
    \label{fig:chi2_comparison}
\end{figure}

The distribution of $\chi^2_{\mathrm{r}}$ values for all OC types for the six targets is presented in Fig. \ref{fig:chi2_comparison}. Equilibrated LL6–7 chondrites emerge as the most plausible analogues for Apophis, Torifune, Eros, and Itokawa, although for the latter, unequilibrated LL3 chondrites also offer a low $\chi_r^2$, for the single available spectrum. Toutatis is clearly best matched by LL3 chondrites. Finally, L chondrites provide the closest match to the Didymos-Dimorphos system spectra, although H chondrites yield fits with comparable reduced $\chi_r^2$ values. In this specific case, we applied a radiative transfer model \citep{Shkuratov1999} that independently confirms L chondrites as the closest analogue. This method has been successfully applied to S-complex spectra \citep{Vernazza_2008, Vernazza_2014, Marsset2024}, and a description of its implementation can be found in those references. 

\subsection{Spectral classification of the S-complex NEO sample}
\label{sec:Distribution_chondrite}

The chondrite matching method was subsequently applied to all other available S-complex NEO spectra ($N=646$). This allowed us to map their distribution in orbital parameter space (semi-major axis, $a$, and inclination, $i$) of OC H-, L-, LL3-, and LL6–7-like NEOs in order to place the results for individual mission targets in a broader context. To search for size-dependent compositional trends as observed between meteorites and NEOs \citep{Vernazza_2008, Broz2024_b}, we further estimated the diameters of these objects. When available, we adopted values from the SsODNet database \citep{SsODNet}. For the remaining objects (approximatively half of the sample), we used their absolute magnitude, $H$, with a fixed geometric albedo, $p_V = 0.25$, a typical value for S-complex NEOs \citep{Delbo_2003,Thomas_2011,Marsset_2022}, to estimate their diameter, $D$, using the standard equation \citep{Bowell_89}

\begin{equation}
    D~=10^{-H/5}\dfrac{1329}{\sqrt{p_V}} ~(\mathrm{km})
.\end{equation}

\begin{table}[!h]
    
    \caption{Diameter distribution of S-complex NEOs grouped by best-match OC type.}
    \label{tab:diameter_distribution}
    \centering
    \begin{tabular}{lccc}   
        \hline
        \vspace{0.2ex} Best match type & $N$ & Median $D$ (km) & 1$\sigma$ interval (km) \\
        
        \hline
        H     & 130 & 0.910 & 0.245 - 2.310 \\
        L     & 89  & 0.693 & 0.108 - 1.450 \\
        LL3   & 53 & 0.370 & 0.081 - 1.179 \\
        LL6–7 & 373 & 0.466 & 0.180 - 1.185 \\

        \hline
    \end{tabular}%
    \tablefoot{$N$ denotes the number of objects, $D$ the median diameter, and we also report the 16–84\% interval.}

\end{table}
Table \ref{tab:diameter_distribution} presents the diameter distributions of each OC-type group.
LL6–7-type objects form the largest group (57.9$\%$). They are followed by H (20.1$\%$), then L (13.8$\%$), and LL3 (8.2$\%$).  This compositional frequency distribution appears broadly consistent with that of kilometre-sized S-complex NEOs from \citet{Broz2024_b}, although our sample contains slightly more H-type and slightly fewer LL6–7-type objects. These minor differences in proportions may arise from (i) the broader diameter range considered here, which is not restricted to kilometre-sized objects, (ii) observational biases inherent to MITHNEOS, and (iii) classification ambiguity near compositional boundaries, where some objects show close $\chi^2_r$ values for two adjacent OC types (see Appendix \ref{app:ambiguity}) that introduce uncertainty in group proportions.

\subsection{Dynamical origin of mission targets}
\label{sec:dyn_origin}
To determine the dynamical origin of our targets, we used probabilistic dynamical maps generated with the model of \citet{Broz2024_a,Broz2024_b}, for the source families of OCs, computed for kilometre-sized particles (NEOMOD). These maps were derived from long-term numerical integrations of test particles released from main-belt asteroid families, including gravitational perturbations from the planets and non-gravitational perturbations due to the Yarkovsky drift. The particles were transported through major resonances, leading to their injection into the near-Earth region. The resulting maps represent probability density distributions in orbital element space, computed from the fraction of simulated particles from each source region that reach a given location in phase space.  
\begin{figure*}[h]
    \centering
    \includegraphics[width=\textwidth]{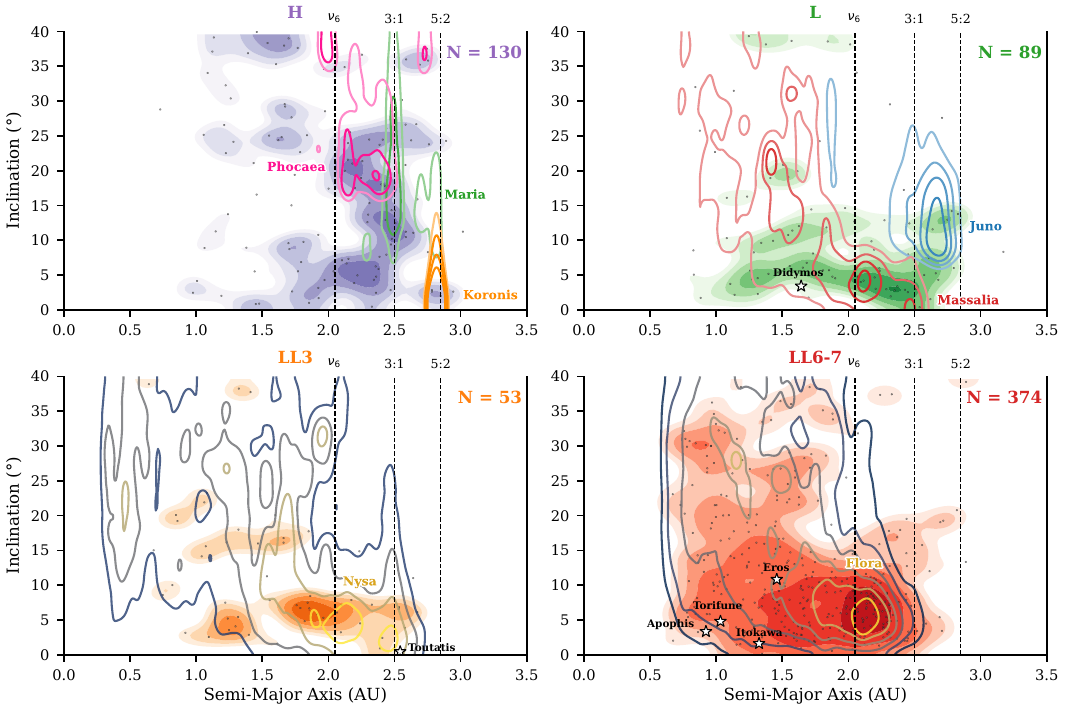}
    \caption{Orbital distribution of S-complex NEOs in the semi-major axis, $a$ (AU), versus inclination, $i$ (°), plane, separated by best-match chondrite type. \textit{Top left}: H. \textit{Top right}: L. \textit{Bottom left}: LL3. \textit{Bottom right}: LL6–7. The background colour (logarithmic scale) shows the local density of objects within each chondrite group; grey dots show individual objects. Overlaid contours show kilometre-scale probability maps from \citet{Broz2024_b} for the main source families of each chondrite type: Phocaea, Maria, and Koronis for H; Juno and Massalia for L; Nysa for LL3; and Flora for LL6–7. White stars mark mission targets: Didymos-Dimorphos system (L panel); Toutatis (LL3 panel); Eros, Itokawa, Torifune, and Apophis (LL6–7 panel). Vertical dashed lines mark the $\nu_6$, 3:1, and 5:2 resonances.}
    \label{fig:proba+density_maps}
\end{figure*}

All LL6–7-like target objects identified by spectral matching (Apophis, Torifune, Itokawa, and Eros) fall within regions of moderate Flora-family probability. Toutatis is located in a region of high Nysa-family probability, near the edge of a high-density area of LL3-like observed NEOs. The Didymos-Dimorphos system lies in a region compatible only with the Massalia-family probability map, excluding a Juno-family origin, the other main contributor to the L-like NEO population \citep{Marsset2024}.
The overall spatial distribution of S-complex NEOs is broadly consistent in shape with the probability maps for L, LL3, and LL6–7 type objects (Fig. \ref{fig:proba+density_maps}). For all OC types, an excess of objects is observed below $a \sim 2$ AU relative to model predictions, likely reflecting the observational biases discussed in Sect. \ref{sec:bias}. For H-like objects especially, the distribution broadly matches the high-density regions predicted for the Phocaea, Maria, and Koronis families but shows a more diffuse component at inclinations below $i \sim 15$°, which was not expected from the simulations.

\subsection{Observational biases}
\label{sec:bias}
\begin{figure}[h]
\includegraphics[width=\columnwidth]{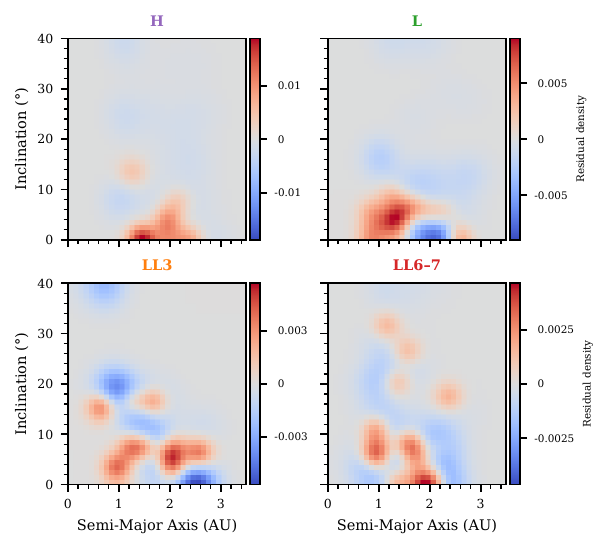}
    \caption{Residual density maps in the semi-major axis, $a$ (AU), versus inclination, $i$ (°), plane, for each OC-like type (H, L, LL3, and LL6–7). Residuals compare the observed density distribution of small NEOs ($H \gtrsim 22$) with the predicted distribution obtained by extrapolating the large-object population ($H < 17$) through the visibility function, $S$. Both distributions are normalised to unit integral over the ($a$, $i$) space before subtraction. Positive values (red) indicate an excess of observed objects relative to the prediction, whereas negative values (blue) indicate a deficit.}
    \label{fig:residus_michael}
\end{figure}

Not all objects are equally likely to be detected: smaller, higher-inclination angle, and greater semi-major axis NEOs are harder to find. Therefore, the observed sample may not uniformly represent the underlying population.
%The observed sample considered is inherently biased: small, high inclination, or distant objects are harder to detect, meaning that observations might not uniformly represent the underlying population. 
To assess whether a potential variation in the compositional distribution with size could result from this bias, we adopted a modelling approach comparing the observed distribution of small objects ($D\leq 100$ m, i.e. $H\gtrsim 22$,\, $N=53$) with a predicted distribution derived from the orbital distribution of larger objects ($D\geq1000$ m, i.e. $H\lesssim 17$,\, $N=175$) of the same OC-like type. The prediction accounts for a visibility function, $S(a,i,H)$, defined as the fraction of time an object would be detectable from Mauna Kea, using an $H,G$ phase function. The eccentricity dimension is not included, as it is averaged within each ($a$, $i$) bin. The size thresholds were chosen to maximise the contrast in size relative to the one-kilometre scale, while maintaining a sufficiently large sample of objects.
For each of the four OC types, we assumed that the observed large-object population provides a reasonable approximation of the intrinsic orbital distribution, as objects of this size should be detectable with sufficient signal-to-noise ratio across most of the accessible orbital space considered. This distribution was smoothed into a probability density map in the ($a$, $i$) orbital space. The predicted density of small objects was then computed as $N_\mathrm{pred}(a,i) = N_\mathrm{obs,\,large}\,(a,i) \times S\,(a,i,H{=}22) \times 10^{\alpha\,(22-17)}$, using the NEO magnitude-distribution slope $\alpha=0.35$ \citep{Bottke_2002}. The observed small-object distribution was similarly smoothed into a density map. Both maps were renormalised such that their integrals over the ($a$, $i$) domain equal one before calculating their difference.
Given the limited sample sizes involved ($N=53$ small objects distributed among the four OC types), the assumption of an unbiased large-object reference population, and the various approximations inherent to the model, this bias-estimation method should be considered qualitative rather than quantitative. Its purpose is to identify the spatial pattern of residual over- and under-densities rather than to provide statistically calibrated estimates of deficits or excesses. The residual maps for each OC-like type are presented in Fig. \ref{fig:residus_michael}. Overall, we identify regions of overabundance of small objects relative to the predicted distribution, primarily at low inclination. This effect is particularly evident for H and L types, with L objects also showing an excess at a low semi-major axis ($a<1.3$ AU). This suggests that, even after accounting for observational biases, the orbital distribution of the smallest objects differs from that of the largest objects.

\subsection{Size-dependent trends}

\begin{figure*}[h]
    \centering
    \includegraphics[width=\textwidth]{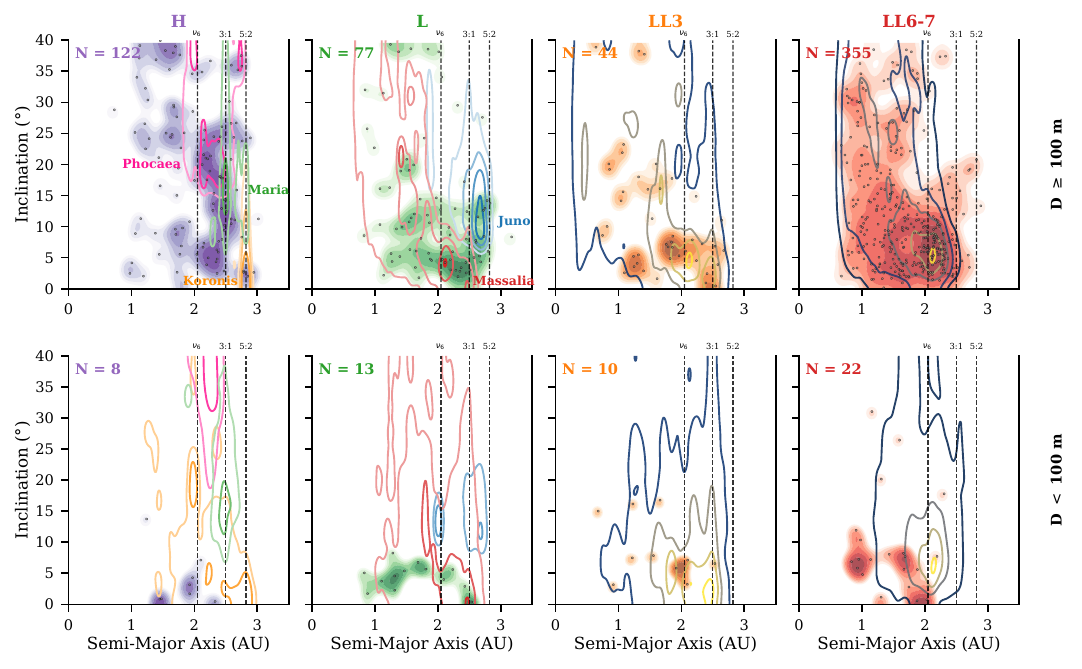}
    \caption{Orbital distribution of S-complex NEOs in the semi-major axis, $a$ (AU), versus inclination, $i$ (°), plane, separated by OC-like analogue (\textit{columns}: H, L, LL3, and LL6–7) and by size (\textit{top row}: $D\geq100$ m; \textit{bottom row}: $D<100$ m). The coloured background and black dots show, respectively, the observed density and individual objects. Overlaid contours represent the density probability distributions from simulations \citep{Broz2024_a,Broz2024_b} of kilometre-sized (top row) and metre-sized (bottom row) objects originating from the main source families: Phocaea, Maria, and Koronis for H-type; Massalia and Juno for L-type; Nysa for LL3-type; and Flora for LL6–7-type objects. The six individually targeted objects (Eros, Itokawa, Toutatis, the Didymos-Dimorphos system, Torifune, and Apophis) are also considered in this figure.}
    \label{fig:map_size_bin}
\end{figure*}

To further investigate this excess and its size dependence, we separated the sample into two size bins: objects with $D \geq 100 $ m and $D<100$ m, as illustrated in Fig. \ref{fig:map_size_bin}. We overlay density contours from simulations of kilometre-sized objects (top row) and metre-sized objects (bottom row). Although the selected size bins do not exactly match the sizes considered in the simulations, they provide an indication of the expected trends: metre-sized objects experience significantly higher drift rates than kilometre-sized objects. This allows us to examine both the evolution of orbital distributions and the relative proportions of each chondrite type with size. 

The dynamical kilometre-scale probability maps are broadly consistent with the observed distribution of the largest objects (top row of Fig. \ref{fig:map_size_bin}), as already discussed in Sect. \ref{sec:dyn_origin}.
However, it is not the case for the small object sample ($D<100$ m, bottom row of Fig.~\ref{fig:map_size_bin}). They are predominantly found at low inclinations and closer to the metre-scale probability maps. For H-like NEOs, the small objects are more consistent with an origin in the Koronis family than in the Phocaea or Maria families, while for L-like NEOs, they are more consistent with a Massalia rather than a Juno origin. For the LL6–7 sample, however, the simulations do not fully reproduce the observations: the cluster at $a\sim1$ AU and $i\sim5$–$10^\circ$ is not reproduced by the models. This clustering may be associated with the dynamically stable ‘Apophis-like’ orbital class characterised by \citet{Broz_2026}, in which NEOs can persist on timescales of tens of millions of years.
This size-dependent shift in orbital distributions is accompanied by a change in compositional proportions.

\begin{figure*}[!h]
    \sidecaption
    \includegraphics[width=12cm]{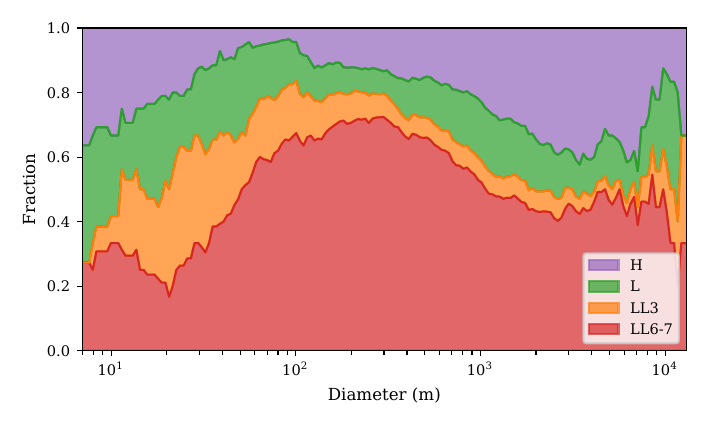}
    \caption{Rolling average proportion of each OC type (H, L, LL3, and LL6–7) as a function of estimated diameter, computed using a sliding window of factor 2 in size. The proportion is normalised to unity at each diameter value.}
    \label{fig:rolling_proportion}
\end{figure*}

Figure \ref{fig:rolling_proportion} shows the rolling average proportion of each chondrite type as a function of object size. LL6–7 objects dominate above $D\sim$ 100 m, and the H fraction increases towards larger sizes. Below $D\sim$ 100 m, the LL6–7 fraction decreases and the H and L fractions increase. LL3 remains a minor contributor throughout most of the size range, with a slight increase below $D\sim$ 100 m. This size-dependent trend is not sensitive to the albedo uncertainty on individual diameter estimates: even under the extreme lower bound of the S-complex albedo range ($p_V=0.10$), the affected diameters would be rescaled by a factor of at most $\sqrt{0.25/0.10}\approx1.6$, which would shift the position of the transition slightly but would not alter its overall shape.
The orbital properties of H- and L-like NEOs with diameters $D\leq$ 100 m are consistent with origins in the Koronis (H) and Massalia (L) families and appear inconsistent with origins in the Phocaea (H) and Juno (L) families \citep{Broz2024_b, Marsset2024}, as is shown in the bottom row in Fig. \ref{fig:map_size_bin}.

\section{Discussion}
\subsection{Individual object analysis}
Overall, the OC-type (H, L, LL3, and LL6–7) classifications derived from spectral matching are consistent with the orbital probability maps of their associated parent families. Hereafter, we discuss our results in the context of previous works, and a summary of these comparisons is presented in Table \ref{tab:mission_summary}.
\begin{itemize}
    \item Eros: Our LL6–7 classification is consistent with previous studies from the NEAR Shoemaker mission. Spectroscopic analyses of data from the Multispectral Imager (MSI) and the Near-Infrared Spectrometer (NIS) favoured L or LL OCs \citep{McFadden2001,Izenberg2003}, while \cite{Lucey2002} identified LL OCs as the best match. Data from the X-ray and gamma-ray spectrometer (XGRS) also supported an L or LL OC composition \citep{Peplowski201636} and ruled out an H chondrite affinity.
    Its orbital properties are consistent with an origin in the Flora family (Fig.~\ref{fig:proba+density_maps}), which represents the primary source of LL6–7 chondrite-like material among main-belt families \citep{Broz2024_b,Marsset2024}.
    
    \item Itokawa: The best match with LL6–7 chondrites is consistent with in situ and laboratory analyses from the {Hayabusa} mission. The NIR spectrometer on board the spacecraft identified an olivine-rich assemblage compatible with LL5–6 chondrites \citep{Abe2006}. Analyses of returned dust particles confirmed a mineralogy and chemical composition consistent with equilibrated LL chondrites \citep{Nakamura2011}, a result also supported by oxygen isotopic measurements, which place Itokawa within the LL or L equilibrated chondrite group \citep{Yurimoto2011}. Independently, ground-based spectroscopic observations prior to the mission also predicted an LL OC composition \citep{Binzel_2001}.
    Its orbital properties are likewise consistent with a Flora family origin. Finally, \citet{Park2015} measured a $^{40}$Ar/$^{39}$Ar age of $1.3 \pm 0.3$ Ga on returned samples, interpreted as an impact or rubble-pile assembly event, which is consistent with the estimated Flora family age of $1.2 \pm 0.2$\,Ga \citep{Broz2024_a}
    and the onset of lunar bombardment ${\sim}1.4$\,Ga ago \citep{Lagain_2026}.
    
    \item Toutatis: Its flyby by the Chang’e 2 spacecraft provided high-resolution optical images of the surface \citep{Huang2013} but no spectroscopic data. Therefore, the only mineralogical constraints are derived from ground-based observations. \citet{Reddy2012} classified Toutatis as an L-chondrite based on NIR band parameter analysis, whereas we obtained a best-fit $\chi^2$ spectral match with LL3 material, with L-chondrite solutions falling outside the uncertainty bounds. 
    Orbital properties are compatible with both Massalia (L) and Nysa (LL3) family members, but show a higher probability density towards the Nysa family. Note that the orbital parameters appear incompatible with a Juno-family origin, which is the main source of kilometre-sized L-type NEOs \citep{Broz2024_b} (Fig. \ref{fig:proba+density_maps}).

    \item Didymos-Dimorphos: The DART mission did not carry an infrared spectrometer; therefore, mineralogical constraints rely on ground-based NIR observations, including pre- and post-impact spectra from \citet{Polishook_2023}. Most spectral analyses suggest best compatibility with L and LL OCs \citep{Polishook_2023,Massa2024}, with a preference for LL. \citet{Rivkin_2023} reaches similar conclusions using combined NIR and mid-infrared (MIR) analyses, although noting that the Band I centre position is more consistent with H or L than LL chondrites. Our $\chi^2_r$ spectral matching favouring L chondrites is therefore consistent with this ambiguity. This L-chondrite affinity implies dynamical compatibility with a Massalia family origin. In particular, the low inclination of the Didymos-Dimorphos system appears incompatible with an origin in the Juno (L) family.
    
    \item Apophis: Both \citet{Binzel_2009} and \citet{Reddy_2018} classify Apophis as LL-like, consistent with our LL6–7 best fit. Its orbital properties are also consistent with a Flora family origin \citep{Broz_2026}.
    
    \item Torifune: Our LL6–7 best match is broadly consistent with the results of \citet{Popescu_2025} and \citet{Bourdelle_2025}, who both infer a composition compatible with L and LL chondrites. Both studies identify the irradiated L6 chondrite Château-Renard as the best spectral analogue, while \citet{Popescu_2025} also finds the LL6 Appley Bridge to be a close second. Repeating our fit using the Château-Renard spectrum, however, we find that it reproduces Torifune's reflectance minimum less well than our LL6–7 mean spectrum, which remains the best overall match. \citet{Bourdelle_2025} further suggest a possible origin in the Lucienne family based on spectral slope matching. However, several of the families they consider, including Flora, are statistically consistent with Torifune's slope within 1$\sigma$. In contrast, Torifune's orbital distribution in our dynamical maps favours a Flora-family origin. Moreover, Lucienne is a relatively small family whose composition has not yet been characterised, making it a less compelling source candidate.
    
\end{itemize}

\begin{table*}[!h]
    \centering
    \renewcommand{\arraystretch}{1.15}
    \caption{Comparison between this-work and literature classification for each individually targeted object.}
    \label{tab:mission_summary}
    \begin{tabular}{l c c c}
        \toprule
        Object & $D$ (m) & This work & Literature \\
               &         & OC-type [family] & OC-type [family] (reference) \\
        \midrule
        Apophis
            & $340 \pm 40$
            & LL6–7 [Flora]
            & LL \citep{Binzel_2009} \\
            & & & LL [Flora] \citep{Reddy_2018} \\
        \addlinespace
        Didymos, Dimorphos
            & $761 \pm 26$, $151 \pm 5$
            & L [Massalia]
            & L or LL \citep{Polishook_2023, Massa2024} \\
            & & & LL, but Band~I position consistent with H or L \citep{Rivkin_2023} \\
        \addlinespace
        Eros
            & $17\,600 \pm 1200$
            & LL6–7 [Flora]
            & L or LL \citep{McFadden2001, Izenberg2003, Peplowski201636}$^\dagger$ \\
            & & & LL \citep{Lucey2002}$^\dagger$ \\
        \addlinespace
        Itokawa
            & $326 \pm 11$
            & LL6–7 [Flora]
            & LL4–6 \citep{Abe2006, Nakamura2011, Park2015}$^\dagger$ \\
            & & & L or LL \citep{Yurimoto2011}$^\dagger$ \\
            & & & LL \citep{Binzel_2001} \\
        \addlinespace
        Toutatis
            & $2725 \pm 194$
            & LL3 [Nysa]
            & L \citep{Reddy2012} \\
        \addlinespace
        Torifune
            & $383 \pm 36$
            & LL6–7 [Flora]
            & equilibrated L or LL \citep{Popescu_2025} \\
            & & & equilibrated L or LL [Lucienne] \citep{Bourdelle_2025} \\
        \bottomrule
    \end{tabular}
    \tablefoot{$\dagger$ indicates results based on orbital or in situ data.}
\end{table*}

The MITHNEOS-based sample is large enough to investigate the compositional and dynamical properties of NEOs as a function of object size. As discussed in Sect.~\ref{sec:bias}, observational bias explains a significant fraction of the differences observed between large and small objects. However, objects below $D\sim 100$ m ($N=53$) appear in excess relative to the bias-corrected predictions (Fig.~\ref{fig:residus_michael}), suggesting that this population is not fully accounted for by detection effects alone. This excess is accompanied by a shift in compositional proportions: below $D\sim 100$ m, the fraction of H- and L-type objects increases at the expense of LL6–7 (Fig.~\ref{fig:rolling_proportion}), tending towards the proportions observed among meteorite falls \citep{Broz2024_b, Marsset2024}. The orbital properties of H- and L-type NEOs with $D \leq 100$ m resemble those of H and L-like bolides observed by camera networks \citep{Marsset2024}. The increasing fraction of H- and L-like NEOs with $D\leq$100 m is consistent with their source families still containing abundant fragments at these sizes, in agreement with their relatively young ages \citep{Broz2024_b}.

\subsection{Future missions prospects}
\begin{figure*}[!h]
    \sidecaption
    \includegraphics[width=12cm]{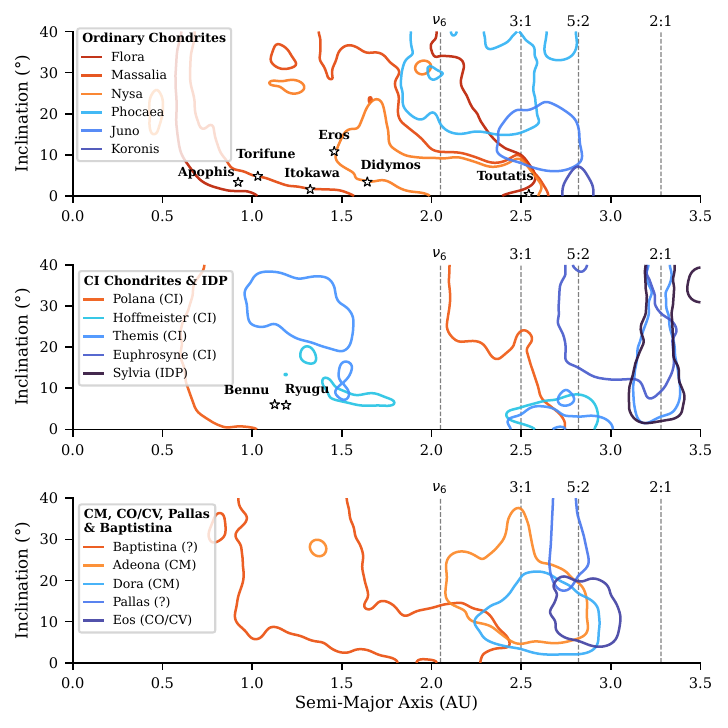}
    \caption{Orbital distribution in the semi-major axis, $a$ (AU), versus inclination, $i$ (°), plane of the main NEO source families from dynamical simulations, separated by compositional class. 
    Top: OC-like source families. Middle: CI chondrite and IDP-like source families. Bottom: CM, CO, and CV chondrite-like source families. Contours indicate the spatial location of the probability density of each source family. Colours range from warm (inner belt, more accessible for space missions) to cool (outer belt, less accessible) with increasing semi-major axis. White stars indicate the orbital positions of space mission targets: the six S-complex NEOs studied in this work, as well as Bennu and Ryugu.}
    \label{fig:future_missions}
\end{figure*}

The orbital distribution of S-complex classified NEOs is broadly consistent with the high-density region of their main OC-like object sources predicted by dynamical simulations (Fig. \ref{fig:proba+density_maps}). Considering this, such simulations appear to provide a reliable guide for identifying orbital regions populated by objects from families other than S-complexes.
They can therefore be exploited to define target orbital ranges for future sample-return missions, in order to sample objects from families not yet visited.
Figure \ref{fig:future_missions} illustrates the orbital distributions of the main NEO source families from dynamical simulations \citep{Broz2024_a,Broz2024_b} divided into composition classes: OC-like objects (H, L, LL3, and LL6–7), CI chondrite and interplanetary dust particle (IDP)-like objects, and CM, CO, and CV chondrite-like objects. The mission targets studied in this work, together with Ryugu and Bennu (JAXA's Hayabusa2 \citep{Yada2022} and NASA's OSIRIS-REx \citep{Lauretta_2024} sample-return missions), are overplotted for reference.
The orbital regions not yet sampled by space missions correspond to higher semi-major axes and higher inclinations, implying higher $\Delta V$ requirements; reaching these targets with traditional rendezvous missions would be challenging.

\section{Conclusions}

In this work, we investigated the source families of six S-complex NEOs visited or to be visited by space missions (Eros, Itokawa, Toutatis, the Didymos-Dimorphos system, Torifune, and Apophis) by combining NIR spectral fitting with space-weathered OC analogues, using an empirical reddening model, and cross-validating with probabilistic dynamical maps of the main NEO family sources.
We identify LL6–7 chondrites as the closest analogues and the Flora family as the most probable source for Apophis, Torifune, Itokawa, and Eros; LL3 chondrites and Nysa for Toutatis; and L chondrites and Massalia for the Didymos-Dimorphos system. Direct measurements (in situ and sample-return analyses), available only for Eros and Itokawa, support our classifications. This agreement suggests that our approach can provide valuable context for interpreting forthcoming mission results: the Didymos-Dimorphos system (\textit{Hera}, 2026), Torifune ({Hayabusa2\#}, 2026), and Apophis (OSIRIS-APEX and RAMSES, 2029).

Applying the same spectral classification to a large MITHNEOS S-complex NEO sample ($N = 651$ objects, with 813 spectra), we find that their orbital distribution is broadly consistent with dynamical predictions for their associated source families for each chondrite type. 
Accounting for observational bias, we observe an overabundance of small objects at low inclinations and small semi-major axes. Below $D\sim$ 100 m, the fraction of H- and L-type objects increases at the expense of LL6–7, approaching the proportions observed among meteorite falls. Our results provide the first direct observational evidence, based on spectroscopic measurements of S-complex NEOs, of the emergence of the source regions of OC meteorites.
Finally, the overall consistency between dynamical simulations and observations suggests that such simulations can guide the selection of future mission targets. Source families other than the ones already explored mostly lie at higher semi-major axes and higher inclinations. Reaching these target regions would require significantly higher $\Delta V$ capability.

\begin{acknowledgements}
P.S., P.V., and L.J. acknowledge support from CNES.
M.B. was supported by GACR grant no. 25-16789S of the Czech Science Foundation.
Part of the data utilised in this publication were obtained and made available by the MITHNEOS MIT-Hawaii Near-Earth Object Spectroscopic Survey. The IRTF is operated by the University of Hawaii under contract 80HQTR19D0030 with the National Aeronautics and Space Administration. The MIT component of this work is supported by NASA grant 80NSSC18K0849. The properties of Solar System objects are from the service SsODNet.ssoCard of the Space Service (SE-OP) of Laboratoire Temps Espace at Paris Observatory through its Solar System Portal (\url{https://ssp.imcce.fr}).
\end{acknowledgements}
\bibliographystyle{aa}
\bibliography{references}

\begin{appendix}
\nolinenumbers
%\twocolumn
\section{Spectra used for mission targets}
\label{app:spectra_table}
\begin{table*}[h]
    \centering
    \caption{Spectra used for each of the six mission targets.}
    \small
    \renewcommand{\arraystretch}{1.3}
    \begin{tabular}{p{3cm}p{9cm}p{3cm}}
        \hline
        Object & MITHNEOS files & Additional spectra \\
        \hline
        Apophis  & a099942.sp117, au2004MN4.sp35, a099942subm.sp35 & \citet{Reddy_2018} \\[3pt]
        Eros     & \raggedright a000433.sp15, a000433.sp101, a000433.sp102, a000433.sp201n1, a000433.sp223, a000433.sp245n2, a000433.9k, a000433.sp17, a000433.sp103 & — \\[3pt]
        Itokawa  & a025143.5k & — \\[3pt]
        Toutatis & \raggedright a004179.sp115n1, a004179.sp03, a004179.sp30, a004179.sp73, a004179.sp74, a004179.sp116 & — \\[3pt]
        Didymos-Dimorphos  & a065803.sp296 & \citet{Polishook_2023} \\[3pt]
        Torifune & a098943.sp320, a098943.sp339 & \citet{Geem_2023}, \citet{DEMEO2009160} \\
        \hline
    \end{tabular}
    \tablefoot{Didymos-Dimorphos additional spectra were acquired before and after the DART impact.}
    \label{tab:spectra_list}
\end{table*}

\section{Mineralogy of L, LL3, and LL6–7 ordinary chondrites}
\label{app:mineralo_distrib}
\subsection{Ordinary chondrite mineralogy distribution }
\begin{figure*}[t]
    \centering
    \includegraphics[width=\textwidth]{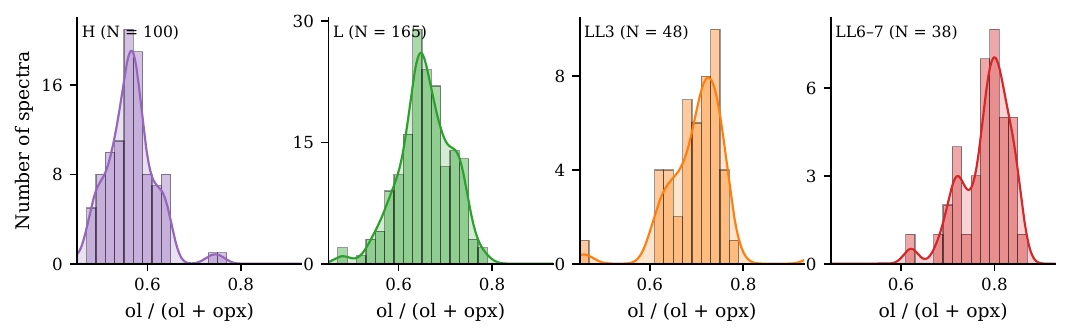}
    \caption{Distribution of the ol/(ol+opx) ratios derived from the Shkuratov model \citep{Shkuratov1999} applied to the individual sample spectra used for the average H, L, LL3, and LL6–7 OC spectra. We observe significant overlap between L and LL3, and between LL3 and LL6–7, indicating that objects near these compositional boundaries cannot be unambiguously classified. This justifies the usage of intermediate mixed spectra, in order to make these in-between spectra absorbed in the neighbouring dominant group (L and LL6–7) rather than LL3. This is a choice motivated by the relative dynamical contributions of the associated source families (see Sect. \ref{sec:Distribution_chondrite}).}
    \label{fig:histo_ratio_ol}
\end{figure*}

\subsection{Intermediate spectral classification justification}
The boundaries between the compositional classes are not intended to represent strict mineralogical divisions between meteorite groups, but rather to reflect the expected compositional distributions of the corresponding asteroid families. LL3 meteorites exhibit a broad mineralogical diversity, with ol/(ol+opx) ratios ranging from $\sim 0.60$ to $0.85$, overlapping both the typical values of L chondrites and those of LL4–LL7 chondrites associated with the Flora family. In contrast, the Nysa family members identified by \citet{marsset2026nysafamilymainsource} occupy a much narrower range, centred around an ol/(ol+opx) value of $\sim$0.73. Therefore, objects with spectra best fitted by an intermediate (L+LL3)/2 composition are classified as L-like rather than Nysa-like, as the full LL3 meteorite range extends beyond the restricted mineralogical domain occupied by the Nysa family.

The boundary between LL3 and LL6–7 is similarly motivated by the relative sizes and compositional distributions of the Nysa and Flora families. Flora is significantly more populous than Nysa \citep{Broz2024_a, Marsset2024}, and its identified members span a broad range of LL petrologic subtypes, from LL4 to LL7, with partial overlap with the LL3 sample considered by \citet{marsset2026nysafamilymainsource}. Consequently, objects with spectra best fitted by an intermediate (LL3+LL6–7)/2 composition are more likely to correspond to Flora members with intermediate petrologic properties than to Nysa-like objects.

\section{Classification ambiguity maps}
\label{app:ambiguity}

For each of the four chondrite classes identified as best-fit types (H, L, LL3, and LL6–7), we assess whether the second-best fit is spectrally close to the best fit.
For each object with best-fit reduced $\chi^2$ value $\tilde{\chi}^2_1$ and second-best fit $\tilde{\chi}^2_2 \geq \tilde{\chi}^2_1$, we define the spectral proximity ratio as follows:

\begin{equation}
  \mathcal{R} \;=\; \frac{\tilde{\chi}^2_2}{\tilde{\chi}^2_1} \;\geq\; 1.
  \label{eq:ratio}
\end{equation}

\begin{figure*}[!h]
  \centering
  \includegraphics[width=\linewidth]{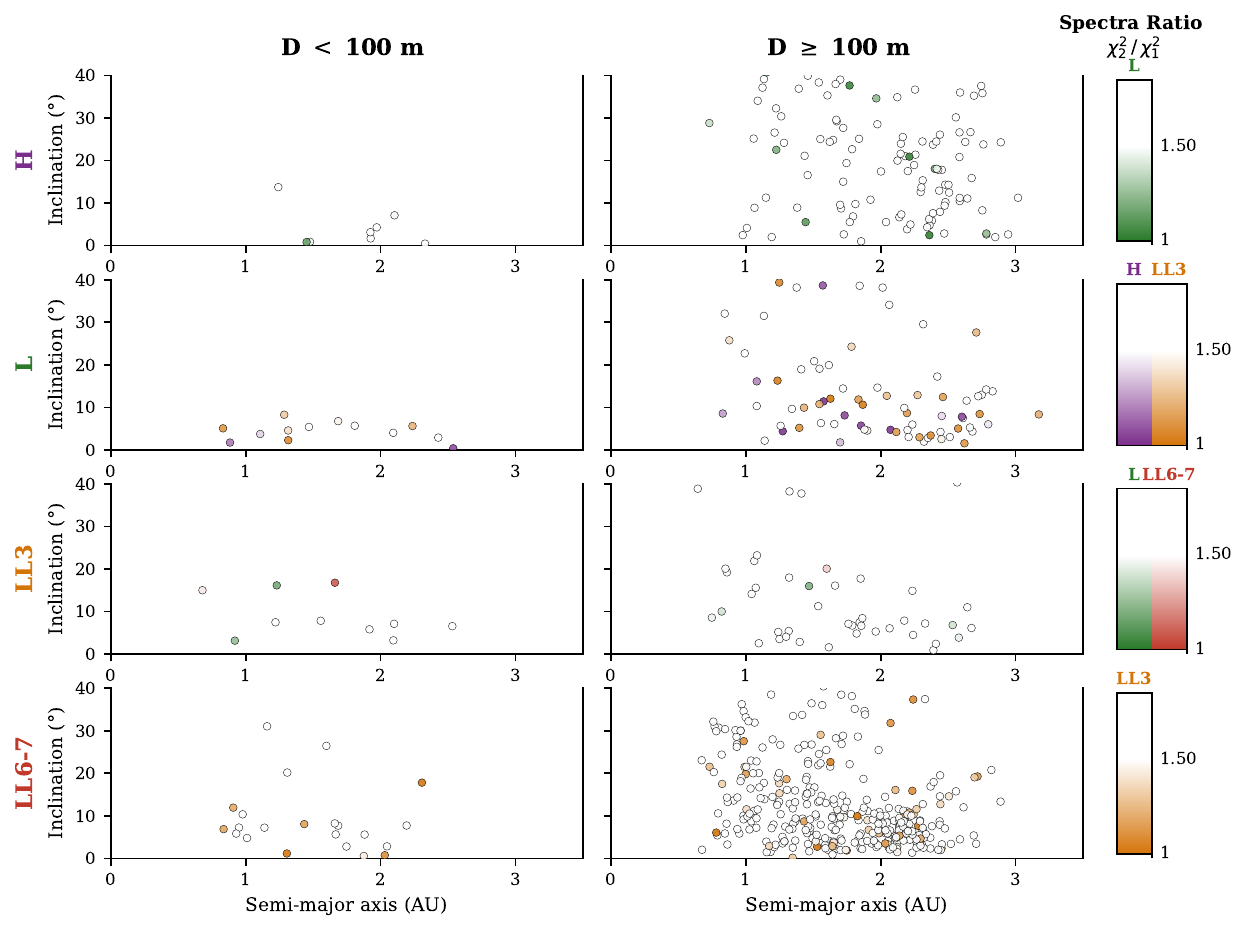}
  \caption{Spectral proximity maps for all four chondrite types (H, L, LL3, and LL6–7, from \textit{top to bottom}). Each row shows the orbital distribution of objects best fitted by a chondrite type, for small bodies ($D< 0.1$ km, \textit{left column}) and larger bodies ($D\geq 0.1$ km, \textit{right column}). 
  The colour of each point represents the spectral proximity ratio $\mathcal{R}$ between the best and the second best fit, with the colour indicating the competitive family.
  Saturated colour indicates nearly identical fits ($\mathcal{R}\approx 1$), while white points ($\mathcal{R}\geq 1.5$) are arbitrarily considered unambiguously classified.}
  \label{fig:ambiguity_all}
\end{figure*}

Overall, Fig. \ref{fig:ambiguity_all} shows that spectral ambiguity is most frequent between adjacent chondrite types along the compositional sequence H–L–LL3–LL6–7. Consequently, L-type and LL3-type objects, which are each bounded by two neighbouring classes, are expected to exhibit the highest degree of ambiguity. However, the fitting strategy described in Sect. \ref{sec:methods} for LL3 objects largely reduces this ambiguity, making them mostly unambiguous. For most NEOs, this residual ambiguity does not appear to follow any clear pattern in orbital properties.
\end{appendix}

\end{document}